\documentclass[onecolumn,showpacs]{revtex4}

\usepackage{graphicx}
\usepackage{dcolumn}
\usepackage{bm}

\input epsf

\begin{document}

\title{\Large A study of higher dimensional inhomogeneous cosmological model}

\author{\bf Subenoy Chakraborty}
\email{subenoyc@yahoo.co.in}
\author{\bf Ujjal Debnath}
\email{ujjaldebnath@yahoo.com}

 \affiliation{Department of
Mathematics, Jadavpur University, Calcutta-32, India.}

\date{\today}

\begin{abstract}
In this paper we present a class of exact inhomogeneous solutions
to Einstein's equations for higher dimensional Szekeres metric
with perfect fluid and a cosmological constant. We also show
particular solutions depending on the choices of various
parameters involved and for dust case. Finally, we examine the
asymptotic behaviour of some of these solutions.
\end{abstract}


\maketitle

\section{\normalsize\bf{Introduction}}
Usually, cosmological solutions to Einstein's field equations are
obtained by imposing symmetries [1] on the space-time. One of the
reasonable assumptions (in an average sense) is the spatial
homogeneity. But when we consider cosmological phenomena over
galactic scale or in smaller scale (detailed structure of the
black body radiation) then we should drop the assumption of
homogeneity i.e., inhomogeneous solutions are useful.\\

Szekeres [2] in 1975 gave a class of inhomogeneous solutions
representing irrotational dust for the metric of the form (known
as Szekeres metric)
$$ds^{2}=dt^{2}-e^{2\alpha}dr^{2}-e^{2\beta}(dx^{2}+dy^{2})$$
Subsequently, the solutions have been extended by Szafron [3] and
Szafron and Wainwright [4] for perfect fluid and they studied
asymptotic behaviour for different choice of the parameters
involved. Later Barrow and Stein-Schabes [5] gave solutions for
dust model with a cosmological constant and showed the validity
of the Cosmic
`no-hair' Conjecture.\\

In this work, we find inhomogeneous solutions for
($n+2$)-dimensional Szekeres space-time with perfect fluid and a
cosmological constant. The paper is organized as follows: The
basic equations are presented in section II while the solutions
have been written in section III. An asymptotic study of
particular solutions are given in section IV. Finally the paper
ends with a short discussion in section V.\\

\section{\normalsize\bf{Basic Equations}}
The metric for the ($n+2$)-dimensional Szekeres space-time is in
the form

\begin{equation}
ds^{2}=dt^{2}-e^{2\alpha}dr^{2}-e^{2\beta}\sum^{n}_{i=1}dx_{i}^{2}
\end{equation}

where $\alpha$ and $\beta$ are functions of all the ($n+2$)
space-time variables i.e., $$\alpha=\alpha(t,r,x_{1},...,x_{n}),~~
\beta=\beta(t,r,x_{1},...,x_{n}).$$ The Einstein equations for
the perfect fluid with a cosmological constant is of the form

\begin{equation}
G_{\mu\nu}=\Lambda g_{ab}+(\rho+p)u_{\mu}u_{\nu}-p g_{\mu\nu}
\end{equation}

where $\rho$ and $p$ are energy density and isotropic pressure
measured by an observer moving with with the fluid, $\Lambda$ is
the cosmological constant and $u_{\mu}$ is the fluid flow four
vector. Since $u=\partial/\partial t$, the flow lines are
geodesics and the contracted Bianchi identities imply that
pressure is a function of $t$ only i.e., $p=p(t)$. As there is no
restriction on the energy density so $\rho$ is in general a
function of all the ($n+2$) variables i.e.,
$\rho=\rho(t,r,x_{1},...,x_{n})$ and hence no equation of state
is imposed.\\

Now from the non-vanishing components of the field equations (2)
for the above metric (1), we have

\begin{eqnarray*}
n\dot{\alpha}\dot{\beta}+\frac{1}{2}n(n-1)\dot{\beta}^{2}-e^{-2\beta}\sum_{i=1}^{n}
\left\{\alpha_{x_{i}}^{2}+\frac{1}{2}(n-1)(n-2)\beta_{x_{i}}^{2}+
(n-2)\alpha_{x_{i}}\beta_{x_{i}}+\alpha_{x_{i}x_{i}}\right.
\end{eqnarray*}
\vspace{-8mm}

\begin{equation}
\left.+(n-1)\beta_{x_{i}x_{i}} \right\}+e^{-2\alpha}
\left\{n\alpha'\beta'-\frac{1}{2}n(n+1)\beta'^{2}
-n\beta''\right\}=\Lambda+\rho
\end{equation}

\begin{eqnarray*}
\frac{1}{2}n(n+1)\dot{\beta}^{2}+n\ddot{\beta}-\frac{1}{2}n(n-1)e^{-2\alpha}\beta'^{2}
-e^{-2\beta}\sum_{i=1}^{n}\left\{\frac{1}{2}(n-1)(n-2)\beta_{x_{i}}^{2}+
(n-1)\beta_{x_{i}x_{i}}\right\}
\end{eqnarray*}
\vspace{-8mm}

\begin{equation}
=\Lambda-p \hspace{-3.4in}
\end{equation}

\begin{eqnarray*}
\dot{\alpha}^{2}+\ddot{\alpha}+(n-1)\dot{\alpha}\dot{\beta}+\frac{1}{2}n(n-1)\dot{\beta}^{2}+
(n-1)\ddot{\beta}+e^{-2\alpha}\left\{(n-1)\alpha'\beta'-\frac{1}{2}n(n-1)
\beta'^{2}-\right.
\end{eqnarray*}
\vspace{-8mm}

\begin{eqnarray*}
\left.(n-1)\beta''\right\}-e^{-2\beta}\sum_{i\ne
j=1}^{n}\left\{\alpha_{x_{j}}^{2}+\frac{1}{2}(n-2)(n-3)\beta_{x_{j}}^{2}+
\alpha_{x_{j}x_{j}}+(n-2)\beta_{x_{j}x_{j}}+(n-3)\alpha_{x_{j}}\beta_{x_{j}}\right\}
\end{eqnarray*}
\vspace{-8mm}

\begin{equation}
-e^{-2\beta}\left\{(n-1)\alpha_{x_{i}}\beta_{x_{i}}+
\frac{1}{2}(n-1)(n-2)\beta_{x_{i}}^{2}\right\}=\Lambda-p
\hspace{-.6in}
\end{equation}

\begin{equation}
\alpha_{x_{j}}(-\alpha_{x_{i}}+\beta_{x_{i}})+\beta_{x_{j}}(\alpha_{x_{i}}+
(n-2)\beta_{x_{i}})-\alpha_{x_{i}x_{j}}-(n-2)\beta_{x_{i}x_{j}}=0,~~~
(i\ne j)
\end{equation}

\begin{equation}
\dot{\alpha}\beta'-\dot{\beta}\beta'-\dot{\beta}'=0
\end{equation}

\begin{equation}
-\dot{\alpha}\alpha_{x_{i}}+\dot{\beta}\beta_{x_{i}}-\dot{\alpha}_{x_{i}}-(n-1)\dot{\beta}_{x_{i}}=0
\end{equation}

\begin{equation}
\alpha_{x_{i}}\beta'-\beta'_{x_{i}}=0
\end{equation}

where dot, dash and subscript stands for partial differentiation
with respect to $t$, $r$ and the corresponding variables
respectively
(e.g., $\beta_{x_{i}}=\frac{\partial\beta}{\partial x_{i}}$) with $i,j=1,2,...,n$.\\

From  equations (7) and (9) after differentiating with respect to
$x_{i}$ and $t$ respectively, we have the integrability condition

\begin{equation}
\dot{\beta}_{x_{i}}\beta'^{2}=0,~~~~i=1,2,...,n
\end{equation}

Thus, if $\beta'\ne 0$ we must have
$\dot{\beta}_{x_{i}}=0,~~i=1,2,...,n$. In the following section we
shall consider the following possibilities $$(i)~ \beta'\ne 0,~~~~
(ii)~ \beta'=0,~ \dot{\beta}_{x_{i}}=0,~~i=1,2,...,n$$ to get
solutions of the field equations.\\

\section{\normalsize\bf{Solutions to the Field Equations}}
In this section we shall solve the field equations (3)-(9) using
the above restrictions separately.\\

{\it Case I}:~~ $\beta'\ne 0$\\

Here due to the restrictions
$\dot{\beta}_{x_{i}}=0,~~i=1,2,...,n$ we have from the field
equations (7) and (9), the form of the metric coefficient as

\begin{equation}
e^{\beta}=R(t,r)~e^{\nu(r,x_{1},...,x_{n})}
\end{equation}

and

\begin{equation}
e^{\alpha}=R'+R~\nu'
\end{equation}

Now substituting these forms for the metric coefficient in
equation (4) we have the differential equations for $R$ and $\nu$
as

\begin{equation}
R\ddot{R}+\frac{1}{2}(n-1)\dot{R}^{2}+\frac{1}{n}(p(t)-\Lambda)R^{2}=\frac{n-1}{2n}f(r)
\end{equation}

and

\begin{equation}
e^{-2\nu}\sum_{i=1}^{n}\left\{(n-2)\nu_{x_{i}}^{2}+2\nu_{x_{i}x_{i}}
\right\}=f(r)-n
\end{equation}

with $f(r)$ as arbitrary function of $r$ alone.\\

Equation (13) can be integrated once to have the first integral

\begin{equation}
\dot{R}^{2}=\frac{2\Lambda}{n(n+1)}R^{2}+f(r)+\frac{F(r)}{R^{n-1}}
-\frac{2}{n}~R^{1-n}\int p(t)R^{n}dR
\end{equation}

where $F(r)$ is another arbitrary function of $r$ (appears due to
integration).\\

Also from (14) the solution for $\nu$ will be

\begin{equation}
e^{-\nu}=A(r)\sum_{i=1}^{n}x_{i}^{2}+\sum_{i=1}^{n}B_{i}(r)x_{i}+C(r)
\end{equation}

with the restriction

\begin{equation}
\sum_{i=1}^{n}B_{i}^{2}-4AC=f(r)-1
\end{equation}

for the arbitrary functions $A(r),~B_{i}(r),~i=1,2,...,n$ and
~$C(r)$.\\

Further, to solve $R$ completely let us consider $p$ as a
polynomial in $t$ as

\begin{equation}
p(t)=p_{0}t^{-a}
\end{equation}

($p_{0}$ and $a$ are positive constants) and we have the general
solution for $R$ as

\begin{equation}R^{\frac{n+1}{2}}=\left\{\begin{array}{lll}

\sqrt{t}\left\{C_{1}J_{\xi}[\frac{2\sqrt{c}}{|a-2|}t^{-\frac{a-2}{2}}]
+C_{2}Y_{\xi}[\frac{2\sqrt{c}}{|a-2|}t^{-\frac{a-2}{2}}]\right\}\\
\\
\sqrt{t}\left\{C_{1}J_{\xi}[\frac{2\sqrt{c}}{|a-2|}t^{-\frac{a-2}{2}}]
+C_{2}J_{-\xi}[\frac{2\sqrt{c}}{|a-2|}t^{-\frac{a-2}{2}}]\right\}\\
\\
C_{1}t^{q_{1}}+C_{2}t^{1-q_{1}}
\end{array}\right.
\end{equation}

according as $\xi$ is an integer, non-integer and $a=2$. Here
$C_{1}$ and $C_{2}$ are arbitrary functions of $r$ and we have
chosen
$$\xi=\frac{1}{a-2},~~c=\frac{(n+1)p_{0}}{2n},~~q_{1}=\frac{1}{2}(1+\sqrt{1-4c}).$$

It is to be noted that to derive the above solution we have
chosen $\Lambda=0=f(r)$. However, for non-zero $\Lambda$ (but
$f(r)=0$) the solution is possible only for $a=0$ and $2$ as

\begin{equation}R^{\frac{n+1}{2}}=\left\{\begin{array}{lll}
C_{1}Cos\{t\sqrt{\frac{n+1}{2n}(p_{0}-\Lambda)}\}+
C_{2}Sin\{t\sqrt{\frac{n+1}{2n}(p_{0}-\Lambda)}\},~~ $when$~~ a=0\\
\\
\sqrt{t}\left\{C_{1}J_{\zeta}[-\frac{i
t\sqrt{\Lambda}}{\sqrt{2}}\sqrt{1+\frac{1}{n}}]+C_{2}Y_{\zeta}[-\frac{i
t\sqrt{\Lambda}}{\sqrt{2}}\sqrt{1+\frac{1}{n}}] \right\}, ~~~
$when$~~ a=2
\end{array}\right.
\end{equation}

with~ $\zeta=\frac{1}{2}\sqrt{1-\frac{2(n+1)p_{0}}{n}}$ .\\

Further, if we consider the dust model (i.e., $p(t)=0$) then the
above solution (19) simplifies to $R^{(n+1)/2}\propto t$. Hence
for the usual 4D (i.e., $n=2$) the scale factor $R$ grows as
$t^{2/3}$ as in the usual Friedmann model.\\

Now, the physical and kinematical parameters have the following
expressions

\begin{equation}
\rho=\frac{n}{2}~\frac{F'+(n+1)F\nu'}{R^{n}(R'+R\nu')}-\frac{p_{0}}{t^{a}}
\end{equation}

\begin{equation}
\theta=\frac{R\dot{R}'+(n+1)R\dot{R}\nu'+n\dot{R}R'}{R(R'+R\nu')}
\end{equation}

\begin{equation}
\sigma^{2}=\frac{n}{8(n+1)(n-1)^{2}}\left[\frac{2R^{n-1}(Rf'-2R'f)+
(n-1)(RF'-(n+1)R'F)}{\dot{R}R^{n}(R'+R\nu')}\right]^{2}
\end{equation}

{\it Case II}:~~ $\beta'=\dot{\beta}_{x_{i}}=0,~~ i=1,2,...,n$\\

In this case from the field equations we have the form of the
metric functions

\begin{equation}
e^{\beta}=R(t)~e^{\nu(x_{1},x_{2},...,x_{n})}
\end{equation}

and

\begin{equation}
e^{\alpha}=R(t)~\eta(r,x_{1},x_{2},...,x_{n})+\mu(t,r)
\end{equation}

Then as before from the field equation (4) we have similar
differential equations in $R$ and $\nu$ as

\begin{equation}
R\ddot{R}+\frac{1}{2}(n-1)\dot{R}^{2}+\frac{1}{n}(p(t)-\Lambda)R^{2}=\frac{n-1}{2n}K
\end{equation}

and

\begin{equation}
e^{-2\nu}\sum_{i=1}^{n}\left\{(n-2)\nu_{x_{i}}^{2}+2\nu_{x_{i}x_{i}}
\right\}=K-n
\end{equation}

with $K$, an arbitrary constant.\\

Here we take the solution for $\nu$ in the form

\begin{equation}
e^{-\nu}=P\sum_{i=1}^{n}x_{i}^{2}+\sum_{i=1}^{n}Q_{i}x_{i}+S
\end{equation}

where the arbitrary constants $P, Q_{i}~(i=1,2,...,n)$ and $S$ are
restricted as before

\begin{equation}
\sum_{i=1}^{n}Q_{i}^{2}-4PS=K-1
\end{equation}

Now to determine the function $\eta$, we have from the field
equation (6)

\begin{equation}
\frac{\partial^{2}(e^{-\nu}\eta)}{\partial x_{i} \partial x_{i}}=0
\end{equation}

and then from the field equation (5) we have the solution

\begin{equation}
e^{-\nu}\eta=u(r)\sum_{i=1}^{n}x_{i}^{2}+\sum_{i=1}^{n}v_{i}(r)x_{i}+w(r)
\end{equation}

with $u(r), v_{i}(r)~(i=1,2,...,n)$ and $w(r)$ as arbitrary functions.\\

Further, to obtain the function $\mu$ we use the following
combination of the field equations namely,

\begin{equation}
\sum_{i=1}^{n}G_{x_{i}}^{x_{i}}-G_{r}^{r}=(n-1)(\Lambda-p(t))
\end{equation}

and the resulting differential equation in $\mu$ is

\begin{equation}
R\ddot{\mu}+(n-1)\dot{R}\dot{\mu}+\mu\left[\ddot{R}+\frac{2}{n}~(p(t)-\Lambda)R\right]=g(r)
\end{equation}

with

\begin{equation}
g(r)=(n-1)\left[2(uS+wP)-\sum_{i=1}^{n}v_{i}Q_{i}\right]
\end{equation}

For explicit solution if we choose $p(t)$ as in the previous case
(see eq.(18)) then the explicit form for $R$ is same as in
equation (19) except here $C_{1}$ and $C_{2}$ are arbitrary
constants and we have chosen $K=\Lambda=0$. Similarly, we have the
same solutions (20) for non-zero $\Lambda$. But we note that the
differential equation (33) is not solvable for any value of $n$.
In fact only for $n=3$ (i.e., for five dimension) we have the
complete solution

\begin{equation}
\mu R=d_{1}t^{q_{2}}+d_{2}t^{1-q_{2}}
\end{equation}

for $a=2$ and
$q_{2}=\frac{1}{2}\left(1+\sqrt{1-\frac{8p_{0}}{3}}\right)$ with
$\Lambda=g(r)=0$.\\

Further, the physical and kinematical parameters have the
expressions as

\begin{equation}
\rho=\frac{2n\dot{\mu}\dot{R}-K}{2\mu
R}+\frac{n}{n-1}\left[\frac{\eta(R^{2}\ddot{\mu}-n\mu
R\ddot{R})-(n-1)\mu^{2}\ddot{R}}{\mu R(\mu+\eta R)}\right]-p(t)
\end{equation}

\begin{equation}
\theta=\frac{R\dot{\mu}+n\mu\dot{R}+(n+1)R\dot{R}\eta}{
R(\mu+\eta R)}
\end{equation}

\begin{equation}
\sigma^{2}=\frac{n}{2(n+1)}\left[\frac{R\dot{\mu}-\dot{R}\mu}{
R(\mu+\eta R)}\right]^{2}
\end{equation}

Finally, for simple dust case we have the solution for
$\Lambda=0$ in terms of hypergeometric function as

\begin{equation}
\sqrt{C_{1}}(t-t_{0})=R^{\frac{n+1}{2}}~~_{2}F_{1}[\frac{1}{2},\frac{n+1}{2n-2},
\frac{3n-1}{2n-2},-\frac{(n+1)^{2}}{4nC_{1}}f(r)R^{n-1}]
\end{equation}

But for non-zero $\Lambda$ we can have solution only for $n=3$ as

\begin{equation}
R^{2}=C_{1}e^{t\sqrt{\frac{2\Lambda}{3}}}+C_{2}e^{-t\sqrt{\frac{2\Lambda}{3}}}-
\frac{f(r)}{\Lambda}
\end{equation}

However, we can have an integral equation from (15) (with $p=0$)
as

\begin{equation}
t-t_{0}=\int\frac{dR}{\sqrt{\frac{2\Lambda}{n(n+1)}R^{2}+f(r)+\frac{F(r)}{R^{n-1}}}}
\end{equation}

and we have the following particular solutions :\\

(i)~~~~$F(r)=0$

$$
R=\sqrt{\frac{(n+1)f(r)}{2\Lambda}}~Sinh\left[(t-t_{0})\sqrt{\frac{2\Lambda}{n(n+1)}}~\right]
$$

(ii)~~~~$f(r)=0$ (i.e., $K=0$ in {\it Case II})

$$
R^{\frac{n+1}{2}}=\sqrt{\frac{(n+1)F(r)}{2\Lambda}}~Sinh\left[(t-t_{0})\sqrt{\frac{(n+1)\Lambda}
{2n}}~\right]
$$

(iii)~~~$\Lambda=0$~~($n=3$)

$$
R^{2}=\frac{1}{3}f(r)(t-t_{0})^{2}-\frac{3F(r)}{f(r)}
$$

(iv)~~~$f(r)=\Lambda=0$

$$
R^{\frac{n+1}{2}}=\frac{n+1}{2}\sqrt{F(r)}~(t-t_{0})
$$

(v)~~~~$f(r)=F(r)=0$

$$
R=e^{(t-t_{0})\sqrt{\frac{2\Lambda}{n(n+1)}}}
$$

Also for the dust $\mu$ has the solution

\begin{equation}
R=\mu=\left[C_{1}e^{t\sqrt{\frac{2\Lambda}{3}}}+C_{2}e^{-t\sqrt{\frac{2\Lambda}{3}}}-
\frac{K}{\Lambda}\right]^{\frac{1}{2}}
\end{equation}

for $n=3$.\\

\section{\normalsize\bf{Asymptotic Behaviour}}
We shall now discuss the asymptotic behaviour of the solutions
presented in the previous section for both perfect fluid and dust
model separately. The co-ordinates vary over the range :
$t_{0}<t<\infty;~ -\infty<r<\infty;~ -\infty<x_{i}<\infty,~
i=1,2,...,n$.\\

\subsection{\normalsize\bf{Perfect fluid model}}
As $p\ge 0,~ p\ne 0$ so we must have $\frac{1}{2}<q_{1}<1$. We
shall first consider the case when $a=2$. For large $t$ ({\it Case
I}~ i.e., $\beta'\ne 0$)
$$
R^{\frac{n+1}{2}}\sim t^{q_{1}}
$$
$$
\rho\sim\frac{n}{2}(F'+(n+1)F\nu')t^{-2q_{1}}
$$
$$
p\sim t^{-2}
$$
$$
\theta\sim t^{-1}
$$
$$
\sigma^{2}\sim t^{-2}
$$

For {\it Case II},~ we have similar behaviour for large $t$
together with $\mu\sim t^{\tilde{q}_{1}}$ where $\tilde{q}_{1}$
is the value of $q_{1}$ for $n=3$ and also we have
$\frac{1}{2}<\tilde{q}_{1}<1$. Thus as $t\rightarrow \infty$,
($p,\rho$) fall off faster compare to ($\theta,\sigma$), while
the scale factor $R$ (and $\mu$) gradually increases with time.
So the model approaches isotropy along fluid world line as
$t\rightarrow \infty$.\\

\subsection{\normalsize\bf{Dust model}}
For the dust case with non-zero $\Lambda$ (and $n=3$) we have for
large $t$,
$$
\mu=R\sim e^{t\sqrt{\frac{\Lambda}{6}}}
$$
$$
\rho=\rho_{0},~~~a~ constant
$$
$$
\theta=\theta_{0},~~~a~ constant
$$
$$
\sigma=\sigma_{0},~~~a~ constant
$$

We note that for {\it Case I},~ $\rho_{0}=\sigma_{0}=0$ while for
{\it Case II}~ we have $\rho_{0}\ne 0,~\sigma_{0}\ne 0$. Thus
universe will behave locally like de-sitter model in {\it Case I}~
though the global geometry will be different. However, in {\it
Case II}~ the universe will not isotropize and for large $t$ the
measure of anisotropy and
expansion scalar become finite constant.\\

\section{\normalsize\bf{Concluding Remarks}}
In this paper, we have found cosmological solutions for
($n+2$)-dimensional Szekeres form of metric with perfect fluid
(or dust) as the matter distribution. We can classify the
solutions in two categories namely, (i) $\beta'\ne 0$ and (ii)
$\beta'=0$. The first set of solutions are known as
quasi-spherical solution while second class of solutions are
termed as cylindrical type of
solutions.\\

However, if we assume the arbitrary functions $A(r)$, $B_{i}(r)$
and $C(r)$ to have the constant values namely
$A(r)=C(r)=\frac{1}{2}$ and $B_{i}(r)=0$ (for all $i=1,2,...,n$)
i.e., if we have chosen the arbitrary function $f(r)$ to be zero
then using the transformation
\begin{eqnarray}\begin{array}{llll}
x_{1}=Sin\theta_{n}Sin\theta_{n-1}...~~ ...
Sin\theta_{2}Cot\frac{1}{2}\theta_{1}\\\\
x_{2}=Cos\theta_{n}Sin\theta_{n-1}...~~ ...
Sin\theta_{2}Cot\frac{1}{2}\theta_{1}\\\\
x_{3}=Cos\theta_{n-1}Sin\theta_{n-2}...~~ ...Sin\theta_{2}Cot\frac{1}{2}\theta_{1}\\\\
....~~ ...~~ ...~~ ...~~ ...~~ ...~~ ...~~ ...\\\\
x_{n-1}=Cos\theta_{3}Sin\theta_{2}Cot\frac{1}{2}\theta_{1}\\\\
x_{n}=Cos\theta_{2}Cot\frac{1}{2}\theta_{1}
\end{array}\nonumber
\end{eqnarray}
the Szekeres metric (1) with the solution (11), (12), (16)
reduces to the ($n+2$)-dimensional spherically symmetric metric
$$
ds^{2}=dt^{2}-R'^{2}dr^{2}-R^{2}d\Omega_{n}^{2}
$$

It is to be noted that if the arbitrary functions $A, B_{i}, C$
depend on $r$, then we can not get the above spherical form of
the space-time. So these arbitrary functions play an important
roll to identify the nature of the space-time.\\

Finally, the study of asymptotic behaviour shows that some of the
solutions will become isotropic at late time while there are
solutions for which shear scalar remains constant throughout the
evolution. Thus Cosmic `no-hair' Conjecture is not valid for all
solutions. This violation of Cosmic `no-hair' Conjecture is not
unusual because the Szekeres metric may not have always
non-positive 3-space curvature scalar. Also in this higher
dimensional Szekeres space-time we have solutions which expands
as de Sitter in some
directions but not in other directions.     \\

{\bf Acknowledgement:}\\

It is a pleasure to thank J. D. Barrow for inspiration to take up
this higher dimensional Szekeres model and for his valuable
comments and discussions in the preliminary draft of the paper.
One of the authors (U.D) is thankful to CSIR (Govt. of India)
for awarding a Senior Research Fellowship.\\

{\bf References:}\\
\\
$[1]$  O. Heckmann, E. Schiicking; in : L. Witten (Ed.): Gravitation, an Introduction
to current research, NewYork, Wiley (1962).\\
$[2]$  P. Szekeres, {\it Commun. Math. Phys.} {\bf 41} 55 (1975).\\
$[3]$  D. A. Szafron, {\it J. Math. Phys.}
{\bf 18} 1673 (1977).\\
$[4]$  D. A. Szafron and J. Wainwright {\it J. Math. Phys.}
{\bf 18} 1668 (1977).\\
$[5]$  J. D. Barrow and J. Stein-Schabes, {\it Phys. Letts.} {\bf 103A} 315 (1984).\\

\end{document}